# High-intensity LINACS: Dynamics, Instabilities and Mitigations

*A. M. Lombardi*
CERN, Geneva, Switzerland

**Abstract**
In this lecture we discuss the intensity limitations in hadron LINACs. First, we will detail what are the two main meanings of intensity limitations as they differ substantially between high power LINACs and high brightness LINACs. Then we will illustrate in detail the building blocks of hadron linacs focusing on the beam dynamics. We will conclude with few examples of intensity limitations and their mitigations.

## 1    Introduction

High intensity LINACs are used in physics laboratories through the world. High-intensity hadron LINACs accelerate a beam of protons or ions to energies of several hundred MeVs to few GeVs and depending on the final energy can be composed of a superconducting section. One of the main uses of high intensity LINACs is to generate intense secondary beams for material and life science studies as for example neutrons at the SNS [1], ESS [2] and JPARC [3]; they can also be used as a source of muons for neutrino factories or muon colliders [4]. For this class of linacs high intensity is reached through high-power, involving high beam currents (from the particle source), high energy (powerful accelerators) and a high duty-cycle. There is another widespread use for LINACs, as for example in the LHC injector chain [5]. These Linacs produce a high intensity proton or ion beam suitable for injection into a synchrotron with a controlled emittance that allows via charge-exchange injection the production of high brightness beam for the luminosity of the collider. In this case the high intensity is reached through the high-quality beam with beam currents of tens of mA in a small emittance (order of 0.2 mm mrad rms normalised).

The main causes of the limitations to high intensity are different in the two cases. In high power linacs the main limitations come from space charge, beam loading, economics, radiation protection, cost of the RF power, cooling and finally activation of the material. In high brightness linacs the limitation comes mainly from the emittance of the particle source and the capability to preserve it during acceleration. This classification is somehow strict as there is overlap between the two regimes, yet it is valid in most of the existing practical cases.

## 2    Building blocks of hadron linacs: radio frequency cavities and magnets

To study the building blocks of a hadron LINAC let's start from the motion equation of a single particle in an electromagnetic field.

Equation (1) describes the motion of a single particle:

$$\frac{d}{dt}(\gamma \frac{d\vec{x}}{dt}) = \frac{q}{m_0} \cdot \left( \vec{E} + \frac{d\vec{x}}{dt} \times \vec{B} \right), \quad (1)$$

where the following definitions hold:

$$\gamma \text{ is the relativistic factor}$$
$$q, m_0 \text{ are the charge and mass of the particle}$$
$$\vec{E}, \vec{B} \text{ are the electric and magnetic field}$$
$$t \text{ is the independent variable time}$$
$$\vec{x} \text{ is the position vector}$$
$$\frac{d\vec{x}}{dt} \text{ is the particle velocity.}$$

Equation (1) says that in order to increase the energy of a beam of particles while keeping them confined in space, we need to provide a longitudinal electric field for acceleration and a transverse force for focusing. To increase the energy of a particle we use a special device called "radio frequency cavity" that transforms electromagnetic energy into energy of the particle via the coupling between the electric field and the charge a particle. To keep the particles bound in a small volume around the direction of propagation we use quadrupole magnets or solenoids. We will speak of the two main building blocks in the next sub-chapters.

## 2.1 Radio Frequency (RF) cavities

Figure 1 shows the sketch of how a radiofrequency cavity work. An RF power source, i.e. a generator of electromagnetic waves of a specified frequency, feeds cavity, i.e. a space enclosed in a metallic boundary which resonates with the frequency of the wave. The electromagnetic wave in the cavity will resonate and sustain a specific field pattern that depends on the geometry of the cavity. When a beam of charged particles passes through the cavity it can increase its speed/energy by taking the stored energy in the cavity via the coupling with the electromagnetic field. In general, there will be an exchange of energy between the particle and the cavity and if the pattern of the field lines and the timing is correct the particle will increase its energy.

The geometry is crucial for the control of the field pattern inside the cavity and to minimise the ohmic losses on the walls while maximise the stored energy. The beam dynamics layout, i.e. the sequence of RF cavities and magnets and their synchronisation, is critical to control the timing between the field and the particle, and to ensure that the beam is kept in the smallest possible volume during acceleration.

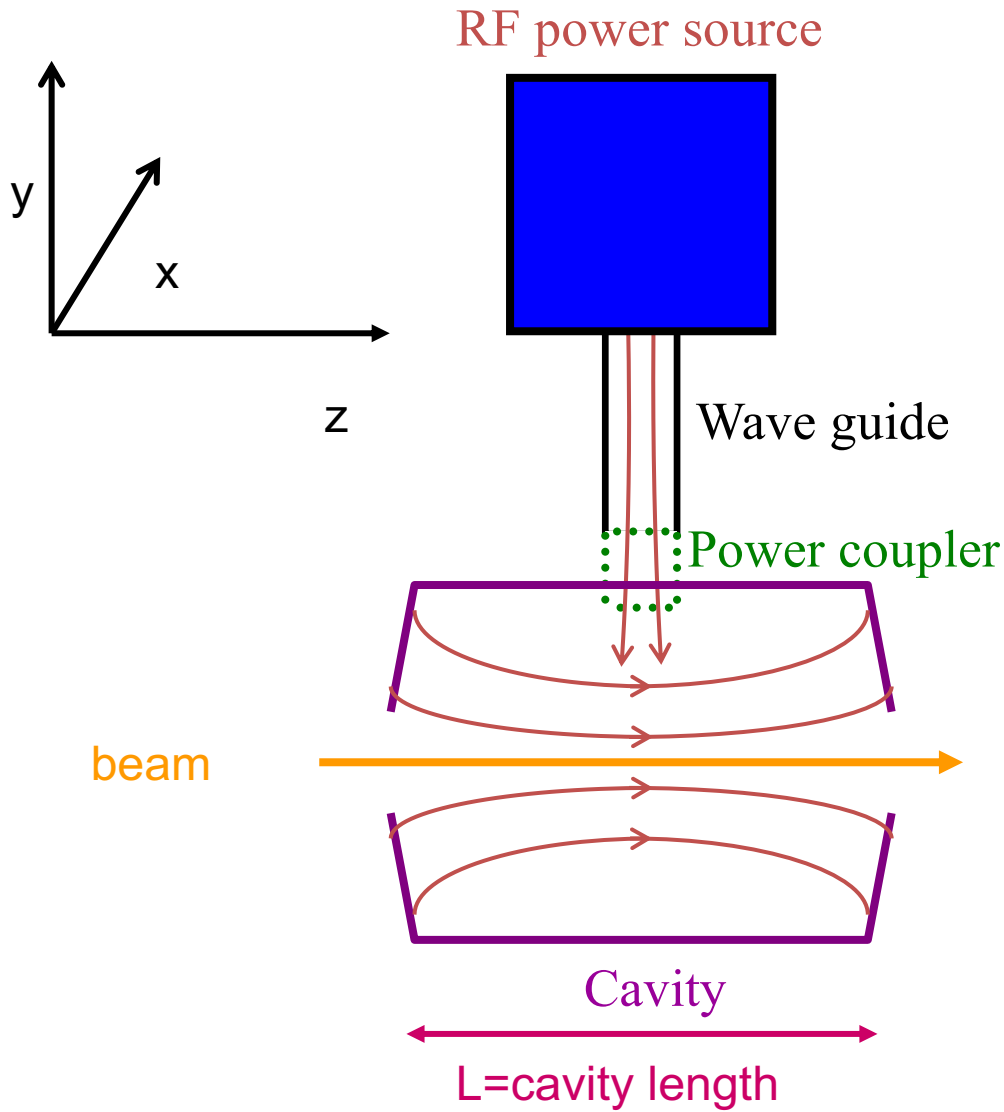

**Fig. 1:** Diagram of a Radio Frequency Cavity

### 2.1.1 Cavity modes and cavity metrics

The electric field in a cavity can be written as the product of two components as in Eq. (2): the first component is a function of space, and the second component is a function of time oscillating between -1 and 1.

$$\mathrm{E}(x, y, z, t) = E(x, y, z) \cdot F(t) \tag{2}$$

The first term describes the patterns of the field or "mode" that can belong to a Transverse Electric mode (TE) or a Transverse Magnetic mode (TM). The word transverse is referred to the direction of propagation i.e. a TE mode will have the transverse component perpendicular to the direction of propagation that in a cavity we can identify as the cavity axis or the axis along which the beam propagates. The most common TE-modes are the dipole mode (TE110) and the quadrupolar mode (TE210) that are used in the Interdigital H structure and the Radio Frequency Quadrupole respectively. The most common TM mode or the fundamental accelerating mode is the TM01 mode. The mode patterns in a cylindrical cavity are shown in Figs. 2 and 3.

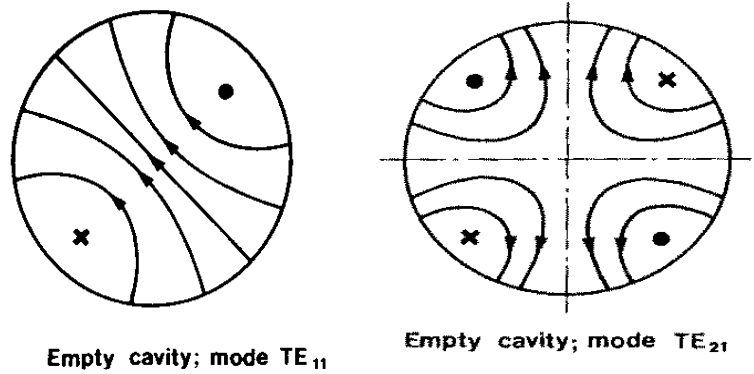

**Fig. 2:** Diagram of the TE11 mode and of the TE12 mode

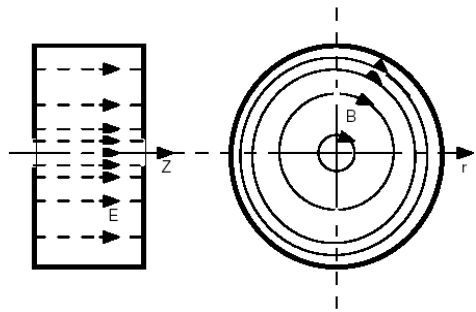

**Fig. 3:** Diagram of the TM01 mode

Resonant RF cavities are fundamental components of particle accelerators. Their role is to generate electromagnetic fields that efficiently transfer energy to charged particle beams, increasing their kinetic energy in a controlled and stable way. The performance of an RF cavity is not described by a single quantity but rather by a set of *cavity metrics* or *parameters* that quantify how effectively the cavity stores energy, dissipates power, and accelerates particles. These metrics depend on the cavity geometry, on the resonant mode, on the properties of the cavity's material, and on the operating frequency. They are an essential tool for accelerator designers.

In the following we introduce the main cavity metrics commonly used in accelerator physics: the average accelerating field, the shunt impedance, the quality factor, the transit time factor, and the effective shunt impedance. Each parameter highlights a different physical aspect of cavity performance, and together they provide a complete picture of how well a cavity is optimized for beam acceleration.

*2.1.1.1   Average Electric Field*

The average electric field, usually denoted by $E_0$ and measured in Volts per meter (V/m), is defined as the spatial average of the electric field along the direction of beam propagation at the moment in time when the RF field reaches its maximum. It is the integral of Eq. (2) along the z axis when F(t)=1, L is the length of the cavity – see Fig. 1. Physically, this quantity represents how much accelerating field is available to the beam inside the cavity.

$$E_0 = \frac{1}{L} \int_0^L E_z(x=0, y=0, z) dz \qquad (3)$$

This parameter is important because the energy gain of a charged particle is directly related to the electric field it experiences along its trajectory. A higher average field generally corresponds to a larger energy gain per cavity, allowing shorter accelerator structures or higher final energies.

The value of $E_0$ depends strongly on the cavity shape and geometry, on the resonant electromagnetic mode excited in the cavity, and on and the operating frequency. Different cavity designs can produce very different field distributions, even at the same frequency. Therefore, the average electric field provides a first, intuitive measure of cavity performance, but it must be complemented by other metrics that account for power dissipation and beam dynamics.

*2.1.1.2 Shunt Impedance*

The shunt impedance, denoted by Z and measured in Ohms per meter (Ω/m), quantifies how efficiently RF power is converted into useful accelerating field. It is defined as the ratio of the square of the average electric field to the RF power dissipated per unit length on the cavity walls:

$$Z = \frac{E_0^2}{P} \cdot L \qquad (4)$$

where P is the power dissipated and L is the cavity length.

From a physical standpoint, the shunt impedance measures how well the cavity concentrates RF power in the region where it contributes to beam acceleration, rather than losing it as heat on the cavity walls. A high shunt impedance means that less RF power is required to achieve a given accelerating field, which is highly desirable for efficient accelerator operation. An important property of the shunt impedance is that it is independent of the absolute field level and of the cavity length. Instead, it depends only on the cavity geometry and the resonant mode. This makes it a powerful figure of merit for comparing different cavity designs on equal footing.

*2.1.1.3 Quality Factor*

The quality factor, or Q factor, is a dimensionless parameter that describes how well a cavity stores electromagnetic energy. It is defined as the ratio between the energy stored in the cavity and the energy lost on the cavity walls during one RF cycle (f in the formula below denotes the frequency):

$$Q = \frac{2 \cdot \pi \cdot f}{P} \cdot U \qquad (5)$$

A high quality-factor indicates that the cavity can store energy for many oscillation cycles before it is dissipated, while a low Q implies significant losses. The quality factor depends on the cavity geometry, which determines the distribution of surface currents, and the surface resistance of the cavity material.

Typical values of Q differ dramatically between technologies: for a frequency of 700MHz e.g. the normal-conducting cavities have Q values of the order of $10^4$, whereas superconducting cavities can reach Q values as high as $10^{10}$, thanks to their extremely low surface resistance.

This enormous difference has major practical consequences. Superconducting cavities require much less RF power to maintain a given field, making them ideal for high-duty-cycle or continuous-wave accelerators, despite their greater technical complexity.

*2.1.1.4  Transit Time Factor*

While the previous parameters describe the electromagnetic properties of the cavity, the **transit time factor (TTF)** directly accounts for the interaction between the RF field and the particle beam. The transit time factor, denoted by $T$, is dimensionless and is defined as the ratio of the actual energy gain of a particle traversing the cavity to the energy gain it would receive if it experienced the maximum field at all times. In reality, the RF field oscillates in time, and particles require a finite time to cross the accelerating gap. As a result, the particle does not always see the peak field. Assuming constant particle velocity (a valid approximation in most cases), the transit time factor can be derived by relating position (z) and time (t) along the cavity axis.

$$T = \frac{\left| \int_0^L E_z(z) e^{-j\left(\frac{2\pi f z}{\beta c}\right)} dz \right|}{\int_0^L E_z(z) dz} \qquad (6)$$

The transit time factor depends on the **particle velocity**, expressed through the relativistic parameter $\beta = v/c$, on the **gap length** $L$, and on the **RF wavelength** $\lambda$.

Importantly, the transit time factor **does not depend on the field amplitude**, but purely on geometry and beam dynamics. If the cavity length is not properly matched to the particle velocity, the particle may even experience partial deceleration instead of acceleration. This highlights the crucial role of the transit time factor in cavity design.

By convention, the transit time factor is usually quoted for particles traveling on the cavity axis. It also depends on transverse coordinates (x, y), in which case we speak of off-axis TTF. The effect of the off-axis TTF is considered in most cases a second order effect and it is not considered at the design stage.

*2.1.1.5  Effective Shunt Impedance*

For accelerator designers, it is often more practical to combine electromagnetic efficiency and beam interaction into a single parameter. This leads to the definition of the **effective shunt impedance**, commonly written as $ZTT$.

The effective shunt impedance incorporates the transit time factor into the shunt impedance, accounting for the fact that not all the available electric field contributes effectively to beam acceleration. It therefore provides a more realistic measure of how well a cavity design is optimized for a given particle velocity.

This parameter is especially useful when comparing cavities intended for different beam energies or particle types. A high effective shunt impedance indicates that the cavity is both electromagnetically efficient and well matched to the beam dynamics, making it a key optimization target in accelerator design.

$$ZTT = \frac{(E_0 T)^2}{P} \cdot L \qquad (7)$$

Cavity metrics play a central role in the design and evaluation of resonant RF cavities for particle accelerators. The average electric field describes the available accelerating strength, the shunt impedance quantifies RF power efficiency, the quality factor measures energy storage and losses, the transit time factor captures the finite interaction time between beam and field, and the effective shunt impedance combines these effects into a beam-relevant figure of merit.

Understanding these parameters and their physical meaning allows accelerator physicists and engineers to design cavities that maximize performance while minimizing power consumption and technical constraints.

*2.1.2 Linac sections and types of RF cavities*

A linear accelerator (LINAC) is composed of several distinct sections, each designed to perform a specific function in the progressive acceleration of charged particle beams. Starting from particle generation and extending to high-energy acceleration, these sections are optimized for different beam energies and relativistic velocities. The overall goal of the LINAC is to efficiently accelerate a beam while preserving beam quality, particularly emittance.

The first stage is the **particle source and DC acceleration**, where particles are generated and initially accelerated to energies ranging from a few keV up to around 100 keV. In this region, the particles are still non-relativistic, with relativistic beta values of at most about 1%. At this early stage, the beam is continuous and has not yet been prepared for radio-frequency (RF) acceleration. A dedicated lecture on this section is part of these proceedings, see Ref. [6].

Following the source, the **pre-injector** section prepares the beam for RF acceleration and provides the first stage of RF acceleration itself. Beam energies in this section typically range from around 0.1 MeV to a few MeV, corresponding to relativistic beta values of approximately 5–10%. This section is crucial because it shapes the beam both transversely and longitudinally so that it can be efficiently captured and accelerated by subsequent RF structures.

The **injector** is a normal-conducting linear accelerator that increases the beam energy to a level suitable for injection into either a circular accelerator (ring) or a superconducting LINAC. Typical energies at the injector exit are of the order of a few hundred MeV, with relativistic beta values around 50%. Finally, **high-energy LINACs**, often made of superconducting cavities, accelerate the beam from roughly 100 MeV up to the GeV energy range. Each of these sections must be carefully matched to the next to minimize beam losses and preserve emittance. In the following subchapters we will investigate the details of few exemplary RF structures for each section.

*2.1.2.1 Radio Frequency Quadrupole*

Within the LINAC chain, the **Radio Frequency Quadrupole (RFQ)** [7] plays a central and unique role. It is the first stage of RF acceleration and is located immediately after the particle source. The RFQ is considered the key component of the pre-injector and is often described as the "missing link" that enabled high-power accelerators.

Before the introduction of RFQs, it was difficult to simultaneously achieve high beam current, small emittance, and efficient acceleration. Typical performance of modern RFQ for high intensity beams include proton beam currents up to 200 mA with normalized emittances of approximately 1 π mm mrad, demonstrating both high intensity and good beam quality.

The RFQ is a linear accelerator that performs **three essential functions simultaneously**: focusing, bunching, and acceleration of a continuous beam of charged particles. A key advantage of the RFQ is that all three processes are carried out using the same RF electromagnetic field, resulting in high efficiency and excellent emittance preservation for high intensity beams.

Transverse focusing in the RFQ is achieved through an **alternating-gradient focusing structure**. The RFQ consists of four electrodes, known as vanes, arranged symmetrically around the beam axis. The beam travels along the axis between the vane tips. The transverse electric field generated by the RF structure provides strong focusing, with a period length equal to $\beta\lambda$, where $\beta$ is the particle velocity relative to the speed of light and $\lambda$ is the RF wavelength. Over half an RF period, the particles travel a distance $\beta\lambda/2$, experiencing alternating focusing forces that confine the beam transversely.

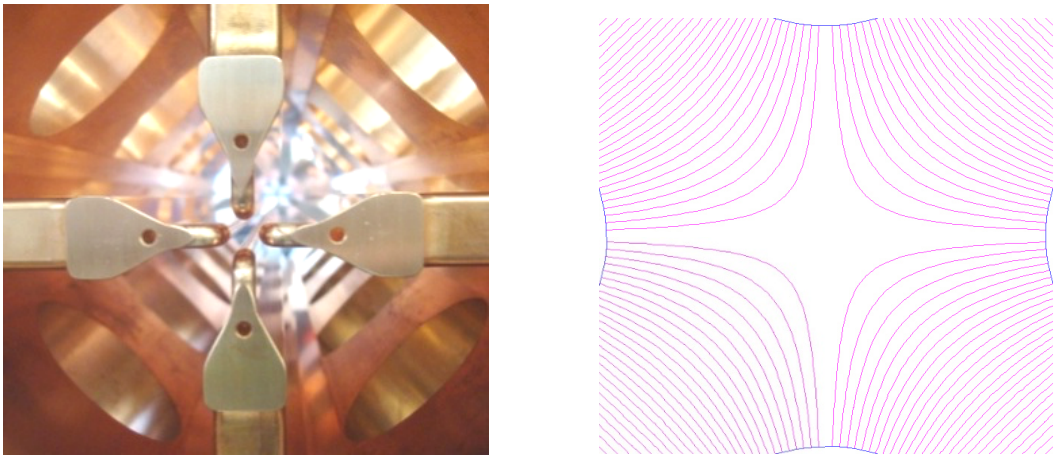

**Fig. 4:** Photo of the transverse profile of an RFQ (beam going into the page) and electric field lines in the beam area. The field lines correspond to the ones of an electrostatic quadrupole. The direction of the field lines changes every half RF period.

Bunching and acceleration within the RFQ is achieved through **longitudinal modulation of the vane tips**. While the RFQ operates in a transverse electric (TE) mode, this modulation introduces a longitudinal electric field component. As a result, particles experience a longitudinal force resulting in bunching and accelerating in addition to focusing. By carefully designing the modulation depth and phase, the RFQ can simultaneously capture, bunch, and accelerate the beam with high efficiency.

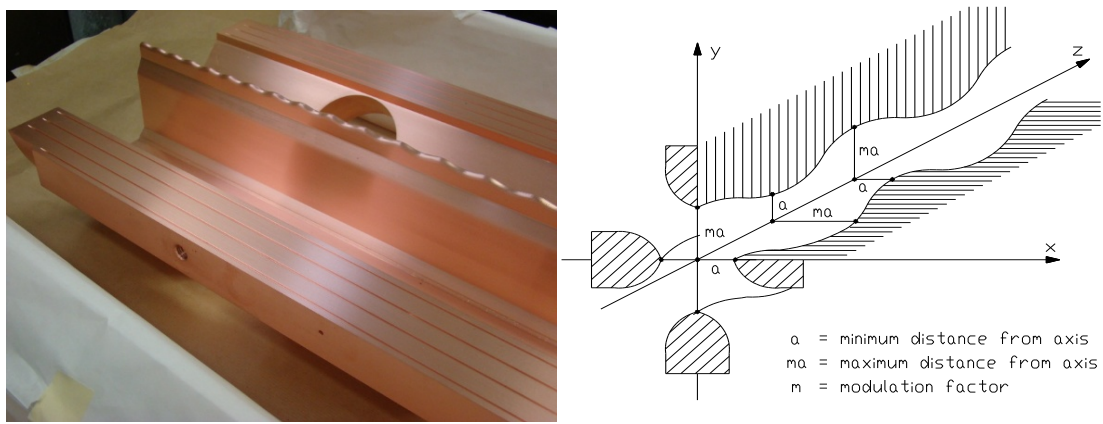

**Fig. 5:** Photo of the longitudinal profile of an RFQ vane and sketch of the period modulation of two adjacent electrodes

In addition to transverse focusing, the RFQ is responsible for **longitudinal bunching** of the beam. Energy transfer from RF fields to particles is not possible for a purely continuous (un-bunched) beam. Therefore, the RFQ must first transform the continuous beam intensity into a bunched beam structure. This is accomplished by smoothly changing the longitudinal velocity profile of the beam without significantly altering its average energy (small modulation factor). As the beam progresses through the RFQ during the bunching process, particles are gradually grouped into RF buckets. Although the average beam current remains the same, the **peak current increases significantly**. A specially tailored bunching section called adiabatic buncher is therefore necessary to preserve the emittance for high intensity beams, resulting in an overall longer RFQ with respect to the equivalent RFQ for low intensity beams [7].

The RFQ represents a major technological advancement in accelerator physics. By providing strong focusing, efficient bunching, and early-stage acceleration in a single structure, it enables the production of high-current, low-emittance beams suitable for injection into more advanced accelerator sections. Its role as the interface between the particle source and the main accelerator makes it indispensable in modern high-power LINACs.

*2.1.2.2 Drift tube linac and other medium energy structures*

The Drift Tube Linac (DTL) is a widely used accelerating structure for protons and heavy ions in the medium-energy range, typically covering particle velocities corresponding to β ≈ 0.04–0.5, which translates to kinetic energies from about 750 keV up to roughly 150 MeV. The DTL operates in the $TM_{010}$ mode (also referred to as the 0-mode), which provides a longitudinal electric field suitable for efficient particle acceleration.

In a DTL, the accelerating cells contain drift tubes that shield the beam when the radiofrequency (RF) electric field is decelerating. This allows particles to experience net acceleration as they traverse successive accelerating gaps. A key feature of the 0-mode field configuration is that removing the walls between cells does not significantly alter the electromagnetic field distribution. This design choice reduces RF currents and leads to a higher shunt impedance, improving overall efficiency.

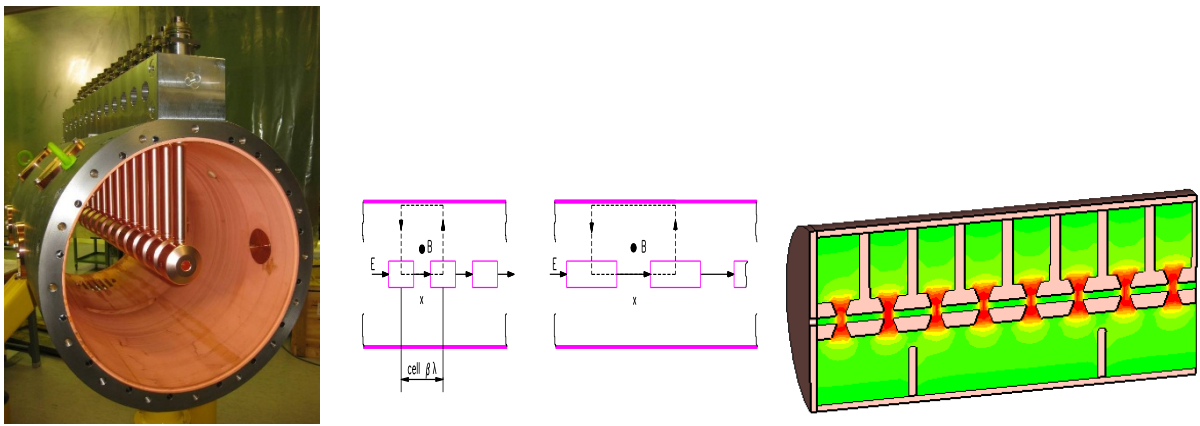

**Fig. 6:** Photo of a DTL model, electrical diagram and electric field distribution.

Synchronism between the RF field and the beam is central to DTL operation. The cell length is proportional to the particle velocity and RF wavelength, ensuring that particles arrive at each accelerating gap at the correct phase of the RF field. This results in a perfectly synchronous structure whereby each drift tube is unique with a different length.

When the beam is sufficiently energetic it is not anymore necessary to maintain perfect synchronism and therefore to simplify construction and reduce cost, DTLs are built as quasi-synchronous structures composed of tanks containing identical cavities rather than individually tailored cells. Multiple tanks are often powered by the same RF source, which further simplifies the RF system. These structures are called CCDTL (cell-coupled DTL) or SCDTL (side-coupled DTL) and are shown in Fig. 5.

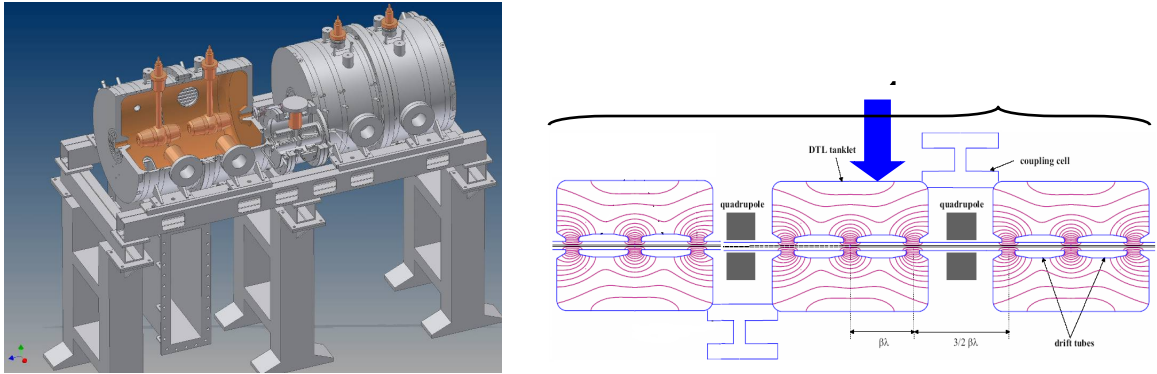

**Fig. 7:** Photo of a DTL model, electrical diagram and electric field distribution.

This simplification introduces the phenomenon of phase slippage. When particles enter a cavity with a velocity lower than the design (geometrical) velocity, they may still be inside the cavity when the RF field changes sign, leading to partial deceleration. As the beam progresses through successive cavities, these small delays accumulate, shifting the phase of the beam with respect to the RF field. The amount of phase slippage is proportional to the number of cavities in a tank and must be carefully optimized. For a given beam velocity, there is a maximum acceptable number of cavities per tank to ensure stable and efficient acceleration.

Focusing in a DTL differs from that in an RFQ (Radio Frequency Quadrupole). While RFQs provide strong continuous transverse focusing using RF fields, DTLs rely on discrete focusing elements, typically quadrupole magnets placed inside the drift tubes or between tanks. This approach balances beam stability with mechanical simplicity and cost effectiveness.

Overall, the DTL remains a cornerstone technology for medium-energy acceleration, combining efficient RF acceleration, manageable beam dynamics, and practical engineering solutions.

For energy higher than 200MeV and certainly for high duty cycle linacs the acceleration is made by superconducting cavities. In order to be efficient and to standardize construction of cavities only few different types of cavities are made for some specific beam velocities and several cavities are grouped in cryostats.

## 2.2 Magnets

After the Radio Frequency Quadrupole (RFQ), transverse beam focusing is provided by magnetic elements, with **solenoids and quadrupoles** being the most widely used. These devices play a crucial role in controlling beam size, stability, and emittance as the particles propagate along the linac.

An infinite long **solenoid** generates a purely axial magnetic field. A solenoid of finite length presents a radial fringe field at the extremities: as charged particles enter the solenoid, the magnetic fringe field induces a transverse velocity component, resulting in a characteristic **rotation of the beam**.

In the central region of the solenoid, where the magnetic field is predominantly longitudinal, the interaction between the particle transverse velocity and the magnetic field leads to a net linear focusing effect that acts symmetrically in both transverse planes. This makes solenoids particularly effective at low energies, where space-charge effects are strong.

In contrast, a **magnetic quadrupole** provides focusing in one transverse plane while simultaneously defocusing the beam in the orthogonal plane. The magnetic field increases linearly with distance from the axis, producing forces that depend on the particle's transverse position. To achieve overall beam confinement, quadrupoles are arranged in alternating sequences of focusing and defocusing elements, such as the well-known **FODO lattice**. In such periodic focusing channels, the beam's four-dimensional phase space is reproduced after each lattice period.

The beam dynamics in a periodic quadrupole channel are governed by **Hill's equation**, which admits periodic solutions when the beam is properly matched [8].

As a result, the Twiss parameters, beta functions, and transverse phase advance repeat from period to period. Focusing lattices are therefore essential tools for maintaining beam quality and mitigating high-intensity effects such as emittance growth, provided that a proper balance between external focusing and collective forces is maintained.

## 2.3 FODO lattice and transverse phase advance, longitudinal phase advance

A linear accelerator is a sequence of cavities interlaced by quadrupole magnets arranged in a FODO sequence, except for the RFQ that provides its own focusing via the electric field. Two important parameters that quantify the combined effect of the RF cavities and the magnet system are the transverse ($\sigma_{0t}$) and longitudinal ($\sigma_{0l}$) phase advance at zero current. In simple term they are a measure of the effective forces in the transverse and longitudinal plane. The phase advance can be measured per focusing period, per meter or per RF gap. Here we choose the convention of using the phase advance per focusing period

$$\sigma_{0t} = \sqrt{\frac{\theta_0^4}{8\pi^2} - \frac{\pi q E_0 T \sin(-\phi)\lambda}{mc^2 \beta \gamma^3}} \qquad (8)$$

with the first term representing the external magnetic force and the second one accounting for the RF defocusing.

$$\theta_0^2 = \frac{qG\lambda^2 N^2 \beta \chi}{m_0 c \gamma}, \qquad (9)$$

where G is the magnetic quadrupole gradient, [T/m] N is the number of magnets in a period (N=2 in a FODO) and χ is the quadrupole filling factor (quadrupole length relative to period length).

$$\sigma_{0l} = \sqrt{\frac{2\pi q E_0 T \lambda \sin(-\phi_s)}{mc^2 \beta \gamma^3}} \qquad (10)$$

# 3 Intensity Limitations in Linear Accelerators: Space Charge Effects and Beam Quality Degradation

In high-intensity linear accelerators, beam dynamics are strongly influenced by collective effects arising from the electromagnetic interaction among charged particles within the beam. **Space charge** plays a dominant role, particularly at low beam energies and high currents. Space charge forces impose fundamental limitations on beam intensity, emittance preservation, and loss control, and therefore directly impact machine performance, reliability, and maintainability.

This section reviews the main physical mechanisms responsible for intensity limitations in linacs, with a focus on space charge effects, their linear and non-linear behaviour, and the consequences for emittance growth and beam halo formation. Strategies commonly adopted to mitigate these effects in modern proton linacs are also discussed.

## 3.1 Space Charge Force: Basic Description

To estimate space charge forces in a linac, the beam is often modelled as a **uniformly charged ellipsoid**, characterized by transverse semi-axes $r_x$ and $r_y$, and a longitudinal semi-axis $r_z$.

Under this assumption, the electric field generated by the beam is linear inside the distribution and vanishes at the center due to symmetry. Contributions from opposite particles cancel out, resulting in zero net force on axis.

$$E_x = \frac{1}{4\pi\varepsilon_0} \frac{3I\lambda}{c\gamma^2} \frac{1-f}{r_x(r_x+r_y)r_z} x$$

$$E_y = \frac{1}{4\pi\varepsilon_0} \frac{3I\lambda}{c\gamma^2} \frac{1-f}{r_y(r_x+r_y)r_z} y$$

$$E_z = \frac{1}{4\pi\varepsilon_0} \frac{3I\lambda}{c} \frac{f}{r_x r_y r_z} z$$

(11)

The space charge force depends on the **beam current** $I$, the beam dimensions, and a geometrical **form factor** $f$, which accounts for the aspect ratio of the ellipsoid [9]. In this idealized case, the force scales linearly with the particle displacement from the beam center, hence it is referred to as a **linear space charge force**.

While this linear approximation is extremely useful for analytical studies, it becomes insufficient when the beam distribution deviates from uniformity or when strong compression occurs, as discussed later.

### 3.1.1 Transverse Phase Advance with Space Charge

The transverse dynamics of a beam in a linac result from the balance of three main forces: the **Quadrupole focusing**, which provides external transverse confinement, the **RF defocusing**, due to accelerating cavities, and the **Space charge defocusing**, arising from particle–particle repulsion

The combined effect of these contributions determines the **transverse phase advance per focusing period**, a key parameter for beam stability, derived from Eq. (8) with the addition of the space charge effects. In the presence of space charge, the phase advance is reduced with respect to the zero-current case. This reduction depends on the beam current $I$, the transverse beam size, and the longitudinal beam dimension $r_z$ with the parameter free-space impedance $Z_0 = 376.73\ \Omega$.

$$\sigma_t = \sqrt{\frac{\theta_0^4}{8\pi^2} - \frac{\pi q E_0 T \sin(-\varphi)\lambda}{mc^2 \beta \gamma^3} - \frac{3Z_0 q I \lambda^3 (1-f)}{8\pi mc^2 \gamma^3 r_x r_y r_z}}$$ (12)

An important consequence is that **bunching or squeezing the beam increases space charge effects**. As the beam is compressed longitudinally, the charge density rises, enhancing the repulsive forces and further depressing the phase advance. This effect is particularly severe at low energies, where relativistic mitigation is weak.

### 3.1.2 *Mitigation Strategies for Space Charge Effects*

Several design strategies are employed to counteract space charge effects in high-intensity linacs:

- **Adiabatic bunching combined with acceleration**, as implemented in RFQs with adiabatic bunchers, to avoid sudden increases in charge density,
- **Fast acceleration**, whenever possible, to reduce the time spent in the low-energy regime,
- **Smooth variation of the phase advance**, avoiding abrupt transitions between strong and weak focusing, and
- **Avoidance of non-linear space charge effects**, which are the main source of irreversible emittance growth.

These principles guide the overall lattice design and strongly influence the choice of focusing strengths and accelerating gradients. These principles translate in several practical design prescriptions commonly applied:

- The **zero-current transverse phase advance** per focusing period should be kept below 90°,
- The phase advance should vary **smoothly along the linac**, independently of beam current, and
- **Resonances must be avoided**, as they can amplify space charge–driven instabilities.

The example in Figs. 8-9 refers to the Drift Tube Linac of LINAC4 [10] where the ratio of phase advance with and without beam current is carefully controlled along the accelerator, ensuring stable beam transport under any beam current conditions.

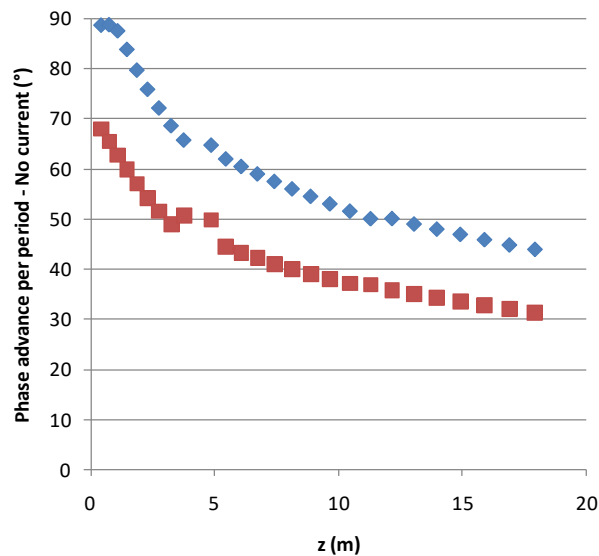

**Fig. 8:** Tranverse (blue) and longitudinal (red) phase advance at zero current along the Drift Tube Linac of LINAC4.

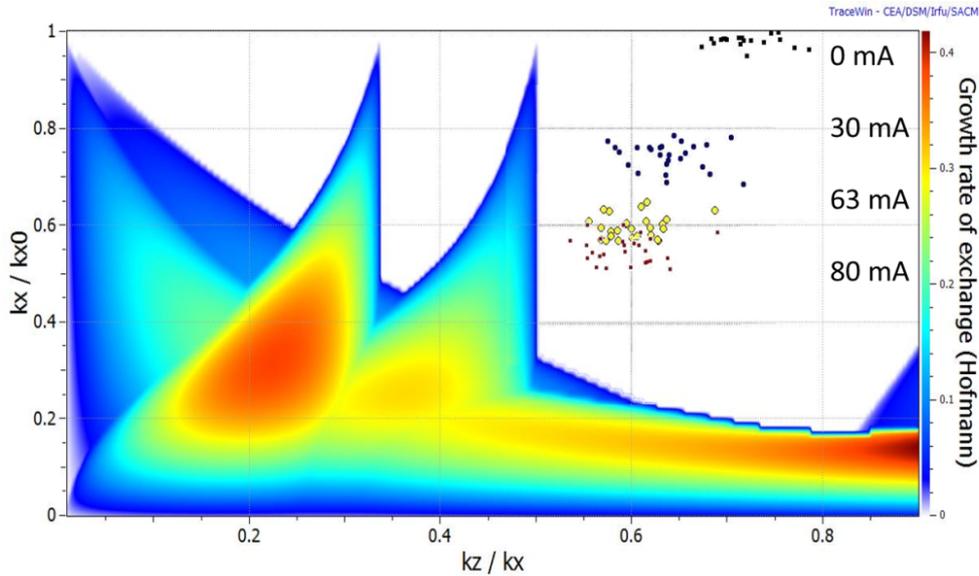

**Fig. 9:** Hofmann resonance plot for the Drift Tube Linac of LINAC4 showing the resonances in the space  σ$_t$/ σ$_{ot}$ vs.  σ$_l$/σ$_t$

### 3.1.3  *Non-Linear Space Charge Effects: Emittance Growth, Filamentation and Halo Formation*

As the beam becomes more compressed in real space, space charge forces become increasingly **non-linear**. In this regime, the force no longer scales linearly with displacement, leading to distortions of the phase-space distribution.

Non-linear space charge effects are particularly harmful because they generate **emittance growth**, which represents an irreversible degradation of beam quality. At low energies, space charge sets a fundamental limit on the **minimum emittance** that can be extracted from an accelerator.

A key mechanism responsible for emittance growth under non-linear forces is **filamentation**. In the absence of space charge, or in the presence of purely linear forces, the rotation frequency in transverse phase space is independent of the oscillation amplitude. With linear space charge, this frequency is reduced but remains amplitude independent.

In contrast, **non-linear space charge introduces an amplitude-dependent rotation velocity**. Different regions of phase space rotate at different speeds, causing the distribution to stretch and fold over time. This process leads to phase-space filamentation and a net increase in emittance. A visualisation of this phenomenon can be seen in Fig. 10.

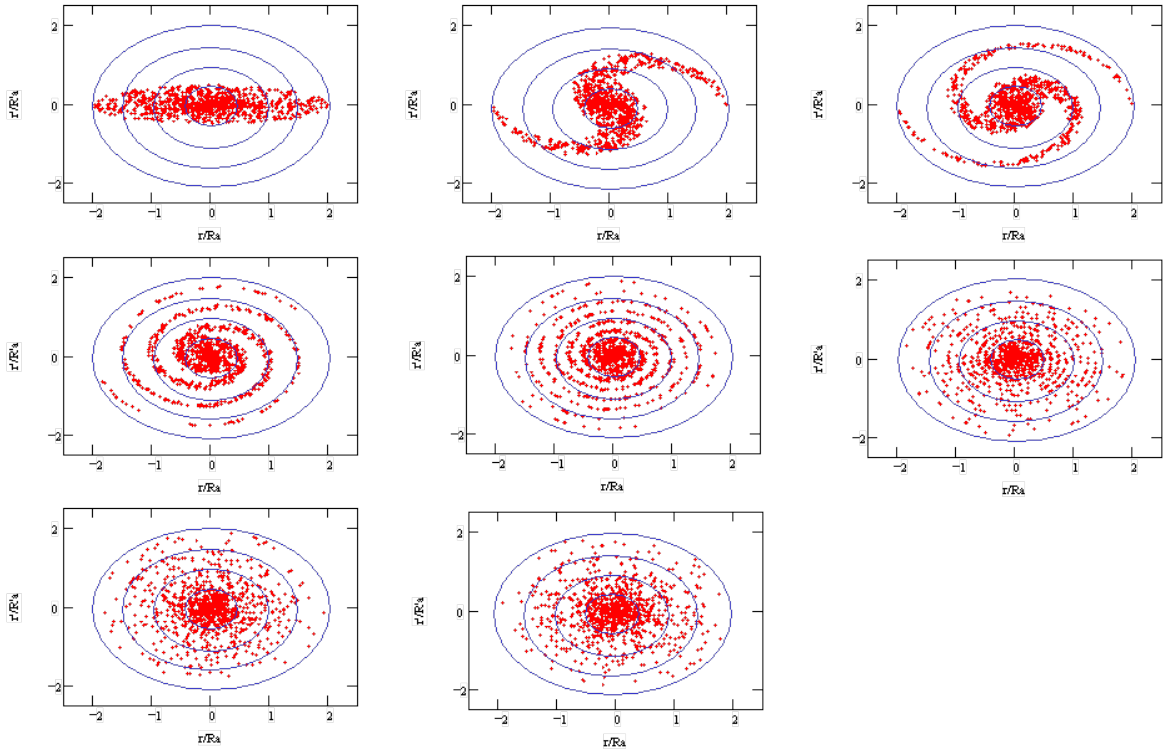

**Fig. 10:** Visualisation of the phenomenon of filamentation. The beam in the transverse phase diffuses into a bigger area due to the variable velocity of rotation of the particles at different radii.

One of the most critical consequences of space charge–induced emittance growth is the formation of a **beam halo**. Halo particles populate large transverse amplitudes and can interact with the accelerator aperture, leading to beam losses.

Beam halo is a major concern in high-power proton linacs because it causes **activation of accelerator components**, making maintenance difficult and time-consuming. For hands-on maintenance, losses must typically be controlled to **below 1 W/m**. For a 2 GeV, 40 mA proton beam, this requirement translates into a calculation accuracy of the order of $10^{-8}$. Accurate modelling of halo dynamics is extremely challenging and lies at the limits of modern computational capabilities, reinforcing the need for conservative and robust design choices.

Space charge effects represent the primary intensity limitation in modern high-power linacs, especially at low energies. While linear space charge can be managed through careful lattice design, non-linear effects inevitably lead to emittance growth, filamentation, and beam halo formation. These phenomena directly impact beam losses, activation, and machine availability.

Effective mitigation relies on adiabatic beam manipulation, fast acceleration, smooth focusing transitions, and conservative phase-advance choices. Together, these strategies form the foundation of reliable, high intensity linac design.

**Acknowledgement**

I wish to thank my colleagues in the Hadron Sources and Linacs section at CERN for the discussions and the material provided.

# References

[1] https://neutrons.ornl.gov/sns last accessed 2 February 2026.

[2] https://ess.eu/ last accessed 2 February 2026.

[3] https://j-parc.jp/c/en/ last accessed 2 February 2026.

[4] Garoby, R. (2010). SPL-based Proton Driver for a nu-Factory at CERN. https://doi.org/10.5170/CERN-2010-003.271.

[5] LIU Technical Design Report CERN-ACC-2014-0337. THPF093 Proceedings of IPAC, Richmond, VA, USA 3918.

[6] D. Faircloth – these proceedings.

[7] A.M. Lombardi, The radiofrequency quadrupole (RFQ), CAS - CERN Accelerator School: small accelerators, pp.201-207, DOI 10.5170/CERN-2006-012.201.

[8] N. Pichoff, Beam dynamics basics in RF linacs, CAS - CERN Accelerator School: small accelerators, pp.145-177, DOI 10.5170/CERN-2006-012.145, https://cds.cern.ch/record/1005047 .

[9] Lapostolle, P. M., Introduction à la théorie des accélérateurs linéaires, ebook: 10.5170/CERN-1987-009.

[10] A.M. Lombardi, G. Bellodi, M. Eshraqi, F. Gerigk, J.-B.Lallement, S.Lanzone, E. Sargsyan, R.Duperrier, D.Uriot, "Beam Dynamics In LINAC4 At CERN", proceeding HB2008, Nashville, Tennessee, August25-29, 2008.